\documentclass[aps,prl,preprint, 10pt,groupedaddress,longbibliography]{revtex4-2}

\usepackage[hidelinks]{hyperref}

\usepackage{amsmath}
\usepackage{mathtools}
\usepackage{longtable}
\usepackage{supertabular,booktabs}
\usepackage{amssymb}
\usepackage{graphicx}
\usepackage{color}
\usepackage{float}
\usepackage{ulem}
\makeatletter
\def\maxwidth{\ifdim\Gin@nat@width>\linewidth\linewidth\else\Gin@nat@width\fi}
\def\maxheight{\ifdim\Gin@nat@height>\textheight\textheight\else\Gin@nat@height\fi}
\makeatother
\setkeys{Gin}{width=\maxwidth,height=\maxheight,keepaspectratio}

\begin{document}

\newenvironment{figurenew}{begin{figure*}}{end{figure*}}
\newenvironment{tablenew}{begin{table*}}{end{table*}}

\title{Liquid water transport model in hydrophilic granular :
Preliminary validation with drying rate of hierarchical granular}
                \author{Hyuga Yasuda}%
                \affiliation{Department of Earth and Space Science,
Osaka University, 1-1 Machikaneyama, Toyonaka 560-0043, Japan}%
                \author{Hiroaki Katsuragi}%
                \affiliation{Department of Earth and Space Science,
Osaka University, 1-1 Machikaneyama, Toyonaka 560-0043, Japan}%
                \author{Makoto Katsura}%
                \email[Makoto
Katsura: ]{mhirai@ess.sci.osaka-u.ac.jp (he/him/his)}
                \affiliation{Department of Earth and Space Science,
Osaka University, 1-1 Machikaneyama, Toyonaka 560-0043, Japan}
            %
            %

    %
    %
      
    \begin{abstract}
    The drying rate profile of granular beds can be divided into the
    constant rate period (CRP), which is characterized by a nearly
    constant drying rate, and the falling rate period (FRP), in which
    the drying rate rapidly decays. In order to explain this behavior
    quantitatively, we proposed a simple one-dimensional power law model
    in which the product of the water permeability and the pressure
    gradient is assumed to be proportional to the cube of the
    saturation. To test this model, we measured the drying rates of
    glass beads and hierarchical granular materials produced by
    sintering and breaking glass beads. Our results and those of
    previous experiments showed consistency with the power law. The
    obtained proportional constant of the experimental power law also
    shows a rough agreement with that estimated from previous studies on
    water permeability and capillary pressure. Drying behavior in FRP
    also agrees with our model in some points. The remnant deviation of
    the model from experimental results may be attributed to the
    inhomogeneity of granular media, which was qualitatively verified.
    \end{abstract}
    
    
    
\maketitle

\section{Introduction}

Many studies have been conducted on the drying rates from porous media.
The experimental results generally indicate a duration with a nearly
constant drying rate (CRP: constant rate period), followed by a rapid
decrease in the drying rate (FRP: falling rate period). In CRP, pore
water is transported to the surface of the porous medium by permeable
flow, and evaporation from the surface continues. However, there is no
unified understanding of the conditions for the transition from CRP to
FRP or the drying rate in FRP. Peat \citep{Palamba2018} and food
products \citep{Srikiatden2005, Roberts2003} exhibit exponential drying
curves in FRP, corresponding to the dehydration process of binding
water. A similar drying rate phenomenon is also possible in powders
owing to the Kelvin effect when the pore diameter is less than a few
nanometers. However, this is not essential for granular beds with large
particle diameters. Exceptionally, this effect has been reported
\citep{Wang2021} in the case of granular beds with diameters of several
hundred microns and very low moisture content(\(\lesssim 1\%\)).

Shokri et al. \citep{Shokri2009, Shokri2010, Shokri2011} saturated the
granular media of sub-millimeter size in a container with dyed water and
measured the drying rate. They used dye deposition to help identify the
boundary between fully saturated, partially saturated, and dry regions.
We cannot ignore the effect of gravity because the depth of the
container exceeds the capillary rise:
\begin{equation}\phantomsection\label{eq:capillaryrise}{
\frac{2\Gamma\cos\theta}{\rho gr},
}\end{equation} where \(\Gamma\), \(\rho\), \(g\), \(r\), and \(\theta\)
are the surface energy of water, the density of water, the acceleration
of gravity, pore radius, and contact angle, respectively. As their
container had a closed bottom, there was always a water-saturated region
at the bottom of the sample. During CRP, the water-containing region
continued to the sample surface because of the water sucked by capillary
forces. During FRP (which they mention as a ``transition period''), a
dry region develops and expands under the surface. As for the period
following FRP, they mentioned it as ``stage-2'', they found that the
depth of the dry region obeys a power law with the system's Bond number
and insisted that the drying rate was controlled by water vapor
diffusion through the dry region. However, they mentioned little about
the transition mechanism from CRP to FRP and the behavior of the drying
rate in FRP.

Although Thiery et al.\citep{Thiery:2017} used bottomed containers, the
effect of gravity was negligible because of their small pore diameters
(mostly sub-micrometers). With MRI, they confirmed the growth of the dry
region in FRP and showed for the first time that the falling rate can be
understood quantitatively by water vapor diffusion through the dry
region. Thiery et al.\citep{Thiery:2017} defined the critical
saturation, \(\langle S \rangle^{*}\), as the spatially averaged water
saturation when the water saturation gradient appears in the wet region
and found that the transition occurs approximately at the same time.
They also showed an empirical formula as below:
\begin{equation}\phantomsection\label{eq:thiery}{
\langle S \rangle^{*}d^{1/4}=1.0\times 10^{-2}\mathrm{m^{1/4}},
}\end{equation}

where \(d\) was originally described as radius; however, we exchanged it
with diameter according to P. Coussot (private communication, June 4,
2024). Here, we define ``transition point'' as the spatially averaged
water saturation at the transition from CRP to FRP. Although the
transition point is expected to be almost the same value as the critical
saturation in granular materials, as discussed above, such a
redefinition is more convenient as the spatial distribution of water
saturation is technically difficult to be observed more than the drying
rate. From here on in this paper, we will consider the critical
saturation to be equivalent to the transition point. They insisted that
the growth of the dry region is controlled by the balance between the
water supply by permeation and evaporative loss. They proposed a model
assuming that water permeability and capillary pressure are proportional
to \(S ^{2}\) and \(S ^{-0.5}\), respectively. Here, the variable \(S\)
means local water saturation. However, the quantitative interpretation
of the measured transition was not successful.

It remains unclear how the proportionality constant shown in
Eq.~\ref{eq:thiery} depends on experimental or sample parameters. In
this study, we will construct a more universal physical model through
further theoretical exploration and perform experimental verification.

\section{Power law model of liquid water transport in hydrophilic
granular media}

\begin{figure}
\centering
\includegraphics[width=1\textwidth,height=\textheight]{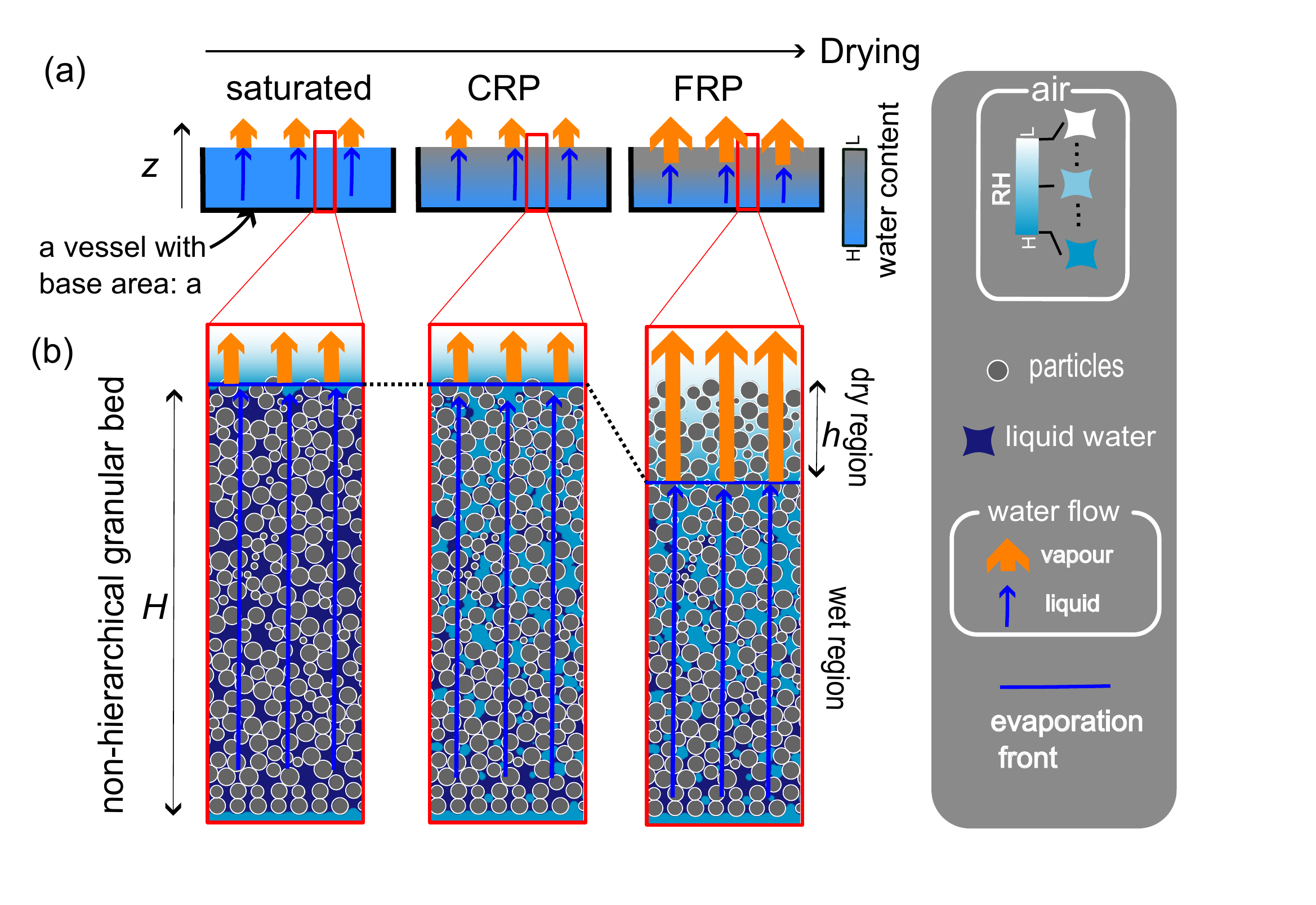}
\caption{Conceptual diagram of a granular bed, (a), and its enlarged
cross-sectional view, (b). Granular was poured in a vessel with the base
area of \(a\). Evaporation creates a gradient in water saturation,
resulting in an upward flow of liquid water. During CRP, air in the pore
is almost 100\% humid, so evaporation only occurs on the top surface of
the granular. After the transition to FRP, a dry region generated at the
top of the granular material recedes to the bottom. As there is no
liquid water here, evaporation only occurs at the boundary with the wet
region. This means that the evaporation front is present on the top
surface in CRP and on the boundary between the two regions in
FRP.}\label{fig:gainen}
\end{figure}

If water transport in the wet region is regarded as the advection of
liquid water driven by capillary pressure gradient and water
permeability\citep{Thiery:2017}, water saturation must satisfy the
following equation: \begin{equation}\phantomsection\label{eq:DL1}{
\begin{aligned}
\frac{\partial S }{\partial t} 
&= \nabla\left (D_\mathrm{L}\nabla S  \right), \\
D_\mathrm{L}&:=\frac{K}{\eta}\frac{dP_\mathrm{c}}{dS },
\end{aligned}
}\end{equation} where \(t\), \(K\), \(\eta\), and \(P_\mathrm{c}\) are
time, permeability of wet region, viscosity of water, and capillary
pressure respectively. This equation is in the form of a diffusion
equation, even though it represents the advection of liquid water. Let
us assume that the apparent diffusion coefficient is proportional to the
\(n\)-th power of water saturation in order to establish the power
dependency of equation Eq.~\ref{eq:thiery}:
\begin{equation}\phantomsection\label{eq:DL2}{
D_\mathrm{L}=D_\mathrm{L}^{0}S ^{n}
}\end{equation} Here, if the sample is a flat granular bed of uniform
thickness as illustrated in Fig.~\ref{fig:gainen} (a) , and the spatial
inhomogeneity of \(D_\mathrm{L}\) can be neglected, Eq.~\ref{eq:DL1}
reduces to a one-dimensional problem in the wet region:
\begin{equation}\phantomsection\label{eq:continuation}{
\frac{\partial S }{\partial t} 
= \frac{\partial}{\partial z}\left (D_\mathrm{L}\frac{\partial S }{\partial z} \right)
}\end{equation}

Here, we set the \(z\)-axis upward from the origin at the bottom as
illustrated in Fig.~\ref{fig:gainen} (a). With the porosity of the
granular, \(\epsilon\), the liquid water flux is given as
\(-\epsilon D_\mathrm{L}\frac{\partial S }{\partial z}\). If the bed has
a closed bottom as in the case of Thiery et al.\citep{Thiery:2017},
liquid water flow is zero at the bottom:
\begin{equation}\phantomsection\label{eq:bottom}{
\left. \epsilon D_\mathrm{L}\frac{\partial S }{\partial z}\right|_{z=0}=0
}\end{equation} Because liquid water flow is limited under the boundary
of dry and wet regions as illustrated in Fig.~\ref{fig:gainen} (b),
integration of Eq.~\ref{eq:continuation} around the boundary leads:
\begin{equation}\phantomsection\label{eq:h}{
S |_{z=H-h}\frac{dh}{dt}=\left.D_\mathrm{L}\frac{\partial S }{\partial z}\right|_{z=H-h}+V,
}\end{equation} where \(h\), \(H\), and \(V\) represents the thickness
of the dry region, thickness of the granular thickness, and the drying
rate as the equivalent velocity of liquid water, respectively. As
illustrated in Fig.~\ref{fig:gainen} (b), both \(h\) and
\(\frac{dh}{dt}\) must be zero during CRP. During FRP, \(S |_{z=H-h}\)
should be negligible due to the decreasing \(D_\mathrm{L}\) at the
receding boundary. As a result, liquid water and evaporative flows are
balanced as follows: \begin{equation}\phantomsection\label{eq:defV}{
\left.D_\mathrm{L}\frac{\partial S }{\partial z}\right|_{z=H-h}=-V
}\end{equation}

According to the definition of \(V\), the relationship with the
spatially averaged water saturation across the granular bed,
\(\langle S\rangle\), is as follows:\\
\begin{equation}\phantomsection\label{eq:Smean}{
\epsilon H \frac{d}{dt}\langle S\rangle=-V, 
}\end{equation}

Thiery et al.~found that water saturation in the wet region has a large
gradient only in a narrow region near the boundary with the dry region;
otherwise, it is almost flat\citep{Thiery:2017}. Based on their
observation, we can presume an approximated solution as follows.

Let \(S_\mathrm{o}\) be the water saturation at the origin (\(z=0\)),
and \(\langle S\rangle_\mathrm{w}\) be water saturation spatially
averaged in the wet region, \(\langle S \rangle H/(H-h)\). Assuming that
the variation of the diffusion coefficient within the flat region is
negligible and can be regarded as a constant value,
\(D_\mathrm{L}^{0}\langle S\rangle_\mathrm{w}^{n}\), the water
saturation can be appropximated as the function, \(F\), defined as
follows: \begin{equation}\phantomsection\label{eq:parabola}{
F(z):=S_\mathrm{o}-\frac{V}{\epsilon H'D_\mathrm{L}^{0}\langle S\rangle_\mathrm{w}^{n}}z^{2},
}\end{equation} where \(H'\) is the height of the flat region and nearly
equal to \(H-h\). Hence, the approximation at the boundary,
\(S_\mathrm{b}\), can be written as:
\begin{equation}\phantomsection\label{eq:Sb}{
S_\mathrm{b}:=F(H-h)\approxeq S_\mathrm{o}-\frac{V(H-h)}{2\epsilon D_\mathrm{L}^{0}\langle S\rangle_\mathrm{w}^{n}}
}\end{equation}

In the CRP far before the transition, saturation in the entire wet
region should be flat; therefore, the diffusion coefficients
corresponding to \(S_\mathrm{o}, S_\mathrm{b}\) and
\(\langle S\rangle_\mathrm{w}\) are \(D_\mathrm{Lo}, D_\mathrm{Lb}\) and
\(\langle D_\mathrm{L} \rangle_\mathrm{w}\) will satisfy the following
equation and fulfill the above condition.
\begin{equation}\phantomsection\label{eq:enough}{
\frac{D_\mathrm{Lo}-D_\mathrm{Lb}}{\langle D_\mathrm{L} \rangle_\mathrm{w}}\ll 1
}\end{equation}

On the other hand, near the transition and in the FRP, a dry region is
considered to be generated and grow because
\hyperref[eq:enough]{Eq.~\ref{eq:enough}} is not satisfied. Then, we can
set the critical condition as follows:
\begin{equation}\phantomsection\label{eq:critical}{
\frac{D_\mathrm{Lo}-D_\mathrm{Lb}}{\langle D_\mathrm{L} \rangle_\mathrm{w}} = m, 
}\end{equation} where \(m\) is a constant not sufficiently smaller than
1. By writing down the critical conditions using
\hyperref[eq:DL2]{Eq.~\ref{eq:DL2}}, the approximate solution can be
written as follows: \begin{equation}\phantomsection\label{eq:kinjikai}{
\begin{aligned}
\langle S\rangle&=\frac{H-h}{H}\left [ \frac{nP}{2m}\right ]^{\frac{1}{n+1}},  \\
P&:=\frac{V(H-h)}{\epsilon D_\mathrm{L}^{0}},
\end{aligned}
}\end{equation} where the definition of the dimensionless quantity \(P\)
has a similar form of the Péclet number; however, the apparent diffusion
coefficient in the denominator represents the advection of liquid water
rather than the diffusion of vapor molecules, and the numerator \(V\)
doesn't represent advection rate but the drying rate that must be
proportional to the vapor molecule diffusion constant. Based on the
analysis so far, by normalizing water saturation with the values of
\(P\) in CRP, and by non-dimensionalizing time and space, we can obtain
the following differential equation with respect to \(\sigma\) and
\(\tilde{h}\): \begin{equation}\phantomsection\label{eq:model1}{
\begin{aligned}
\frac{\partial \sigma}{\partial \tilde{t}}& = \frac{\partial}{\partial \tilde{z}}\left ( \sigma^{n}\frac{\partial}{\partial \tilde{z}}\sigma\right), & (0 \le \tilde{z} \le 1-\tilde{h}) \\
\sigma(\tilde{t}, \tilde{z})&:= P_{0}^{\frac{-1}{n+1}}S , \\
P_{0}&:=\frac{V_{0}H}{\epsilon D_\mathrm{L}^{0}},\\
\tilde{z}&:=z/H, \\
\tilde{t}&:=t/\tau_{0}, \\
\tau_{0}&:=\frac{\epsilon H}{V_{0}}P_{0}^{\frac{1}{n+1}},\\
\tilde{h}&:= h/H,
\end{aligned}
}\end{equation} where \(V_{0}\) is the value of \(V\) in CRP. The
boundary conditions are as follows:
\begin{equation}\phantomsection\label{eq:model2}{
\begin{aligned}
\frac{\partial \sigma}{\partial \tilde{z}}&=0 & (\tilde{z}=0), \\
\frac{\partial \sigma}{\partial \tilde{z}}&=-\frac{V}{V_{0}}\sigma^{-n} & (\tilde{z}=1-\tilde{h}). 
\end{aligned}
}\end{equation}

Consider the solution (\(\sigma, \tilde{h}\)) of
\hyperref[eq:model1]{Eq.~\ref{eq:model1}} and
\hyperref[eq:model2]{Eq.~\ref{eq:model2}} with the uniform distribution
of \(\sigma\) as initial condition. If the initial value of \(\sigma\)
is sufficiently large, \(\sigma\) should immediately transition to a
quadratic function corresponding to Eq.~\ref{eq:parabola} and maintain
this distribution shape until near the transition. If \(\tilde{h}\)
turns positive at the time \(\tilde{t}=\tilde{t}^{*}\), this corresponds
to the transition to the FRP, so the transition point is given by the
following formula: \begin{equation}\phantomsection\label{eq:sol1}{
\begin{aligned}
\langle S \rangle^{*}&=\sigma^{*}P_{0}^{\frac{1}{n+1}}, \\
\sigma^{*}&:=\int_{0}^{1} \sigma(\tilde{t}^{*} , \tilde{z}) d\tilde{z}.
\end{aligned}
}\end{equation} Relative change in drying rates in FRP is given as:
\begin{equation}{
\begin{aligned}
\frac{V}{V_{0}}&=\frac{\langle \sigma \rangle'(\tilde{t})}{\langle \sigma \rangle'(\tilde{t}^{*})}\\
\langle \sigma \rangle'(\tilde{t})&:=\frac{d\langle \sigma \rangle(\tilde{t})}{d\tilde{t}}\frac{1}{\tau_{0}}\\
\langle \sigma \rangle(\tilde{t})&:=\int_{0}^{1} \sigma(\tilde{t} \ge \tilde{t}^{*} , \tilde{z}) d\tilde{z}
\end{aligned}
}\end{equation} We can also rewrite the approximate solution
\hyperref[eq:kinjikai]{Eq.~\ref{eq:kinjikai}} in the non-dimensionalized
variable \(\sigma, \tilde{h}\) as follows:
\begin{equation}\phantomsection\label{eq:kinjikai2}{
\frac{\langle S\rangle}{\langle S \rangle^{*}}=\frac{\langle \sigma \rangle}{\sigma^{*}}=(1-\tilde{h})^{\frac{n+2}{n+1}}\left(\frac{V}{V_\mathrm{0}}\right)^{\frac{1}{n+1}}.
}\end{equation}

As discussed for non-hierarchical granular beds, the Kelvin effect of
the microscopic pores is not significant because of their considerable
grain size \citep{Thiery:2017} and the observed water saturation
\citep{Wang2021}. Therefore, it is reasonable to assume that in the FRP,
a dry region expands on the granular bed surfaces; the falling rate is
due to the diffusion of vapor through the growing thickness of the dry
region (\(h\)). Then, we can express the drying rate in the same way as
in the case of a non-hierarchical granular bed \citep{Thiery:2017}:
\begin{equation}\phantomsection\label{eq:V}{
\begin{aligned}
\frac{V}{V_\mathrm{0}} &=  \frac{ 1}{ 1 + N \tilde{h}} \\
N & :=\frac{H\tau_\mathrm{g}}{\delta \epsilon}
\end{aligned}
}\end{equation} where \(\epsilon\) is the porosity of the granular bed,
\(\tau_\mathrm{g}\) is the diffusion tortuosity in the dry region, and
\(\delta\) is the thickness of the boundary layer in air neighboring the
water surface exposed at the sample surface..

To compare the three equations obtained above, Eq.~\ref{eq:defV},
Eq.~\ref{eq:kinjikai2}, and Eq.~\ref{eq:V}, with experimental results in
FRP, let us eliminate \(\tilde{h}\). First, Eq.~\ref{eq:kinjikai2} and
Eq.~\ref{eq:V} can be jointly constructed as follows.
\begin{equation}\phantomsection\label{eq:kinjikai3}{
\frac{\langle S\rangle}{\langle S \rangle^{*}}=\frac{\langle \sigma \rangle}{\sigma^{*}}=(1-\tilde{h})^{\frac{n+2}{n+1}}(1+N\tilde{h})^{-\frac{1}{n+1}}
}\end{equation} This formula can be regarded as a function that gives a
numerical solution to \(\tilde{h}\) given the ratio of saturation to
transition point (\(\langle S\rangle/\langle S \rangle^{*}\)) in the
FRP. In addition, by jointly establishing Eq.~\ref{eq:defV} and
Eq.~\ref{eq:V}, we can obtain the following equation giving the time
derivative of \(\langle \sigma \rangle\):
\begin{equation}\phantomsection\label{eq:kinjikai4}{
\frac{\epsilon H}{V_{0}}\frac{d}{dt}\langle S\rangle=\frac{d}{d\tilde{t}}\langle \sigma \rangle=-\frac{1}{1+N\tilde{h}}.
}\end{equation} Generally, it is easy to obtain numerical solutions from
a function that gives such a time derivative, and we used the solve\_ivp
function of scipy \citep{2020SciPy-NMeth} in this study. In addition, a
plot showing the relationship between \(\langle S\rangle\) and \(V\) can
be easily obtained by giving common numerical values to \(\tilde{h}\) in
both Eq.~\ref{eq:V} and Eq.~\ref{eq:kinjikai4}.

Based on the Young-Laplace equaition, capillary pressure,
\(P_\mathrm{c}\), must be proportional to the surface energy of water,
\(\Gamma\), and inversely proportional to the microscopic pore radius,
which is proportional to \(d\). where \(\widetilde{P}_\mathrm{c}\) is a
non-dimensional function of \(S\) and is independent of \(d\).\\
Permeability, \(K\), must be proportional to the square of the
microscopic pore radius.

Therefore, \(D_\mathrm{L}\) can be rewritten as follows:
\begin{equation}\phantomsection\label{eq:tilde_D}{
D_\mathrm{L}=\widetilde{D_\mathrm{L}^{0}}  \frac{\Gamma}{\eta}S ^{n} d,
}\end{equation} where \(\widetilde{D_\mathrm{L}^{0}}\) is a
non-dimensional constant that does not depend on \(d\). Substituting
this into the solution \hyperref[eq:sol1]{Eq.~\ref{eq:sol1}} of the
power-law model gives the following:
\begin{equation}\phantomsection\label{eq:trans_power}{
\begin{aligned}
\langle S \rangle^{*}&=\widetilde{P_{0}}^{\frac{1}{n+1}}\widetilde{D_\mathrm{L}^{0}}^{-\frac{1}{n+1}}\sigma^{*} \\
\widetilde{P_{0}}&:= \frac{\eta V_{0}H}{\epsilon \Gamma d}
\end{aligned}
}\end{equation}

Using Eq.~\ref{eq:model1}, we can describe the spatially averaged water
saturation in FRP as: \begin{equation}{
\langle S\rangle=\mathrm{P_0}^{-\frac{1}{n+1}}
\langle \sigma \rangle(\tilde{t})
}\end{equation} We set the approximation function of
\(\langle \sigma \rangle\) in an exponential form as:
\begin{equation}\phantomsection\label{eq:exponential2}{
\langle \sigma \rangle \sim A' \exp(-\tilde{t}/\tilde{\tau})+C'.
}\end{equation} The measured drying time constant, \(\tau\), is
expressed as follows: \begin{equation}\phantomsection\label{eq:tau}{
\tau = \tilde{\tau}\frac{\epsilon H}{V_{0}} \mathrm{P_0}^{\frac{1}{n+1}}
}\end{equation} From Eq.~\ref{eq:tau} and Eq.~\ref{eq:sol1}, we can
obtain a proportionality as:
\begin{equation}\phantomsection\label{eq:prop}{
\tau \frac{V_{0}}{\epsilon H}: \langle S \rangle^{*} = \tilde{\tau}: \sigma^{*}=const.
}\end{equation}

In the following sections, we will evaluate this power-law model by
adding new experimental data, comparing them with this model, and
considering appropriate values for \(n\).

\section{Experiment}

\subsection{Sample preparation}

\begin{table}
$$\nonumber
\begin{aligned}
& \text{TABLE I.} \\
& \begin{array}{crrrr}
\text{sample name}  & d(\mu\text{m}) & D(\mu\text{m}) & \epsilon & \epsilon_\mathrm{m} \\
\hline
\text{non-hi(5)}   &  5 &         - & 0.51 & - \\
\text{non-hi(18)}  & 18 &         - & 0.46 & - \\
\text{non-hi(400)} & 400 &        - & 0.40 & - \\ 
\text{hi-d18(XS)}  & 18 &    74-250 & 0.71 & 0.25 \\
\text{hi-d18(S)}   & 18 &   250-840 & 0.72 & 0.24 \\
\text{hi-d18(M)}   & 18 &  840-2000 & 0.69 & 0.27 \\
\text{hi-d18(L)}   & 18 & 2000-4760 & 0.64 & 0.31 \\
& \end{array}
\end{aligned}
$$
\end{table}

\begin{figure}
\centering
\includegraphics[width=0.5\textwidth,height=\textheight]{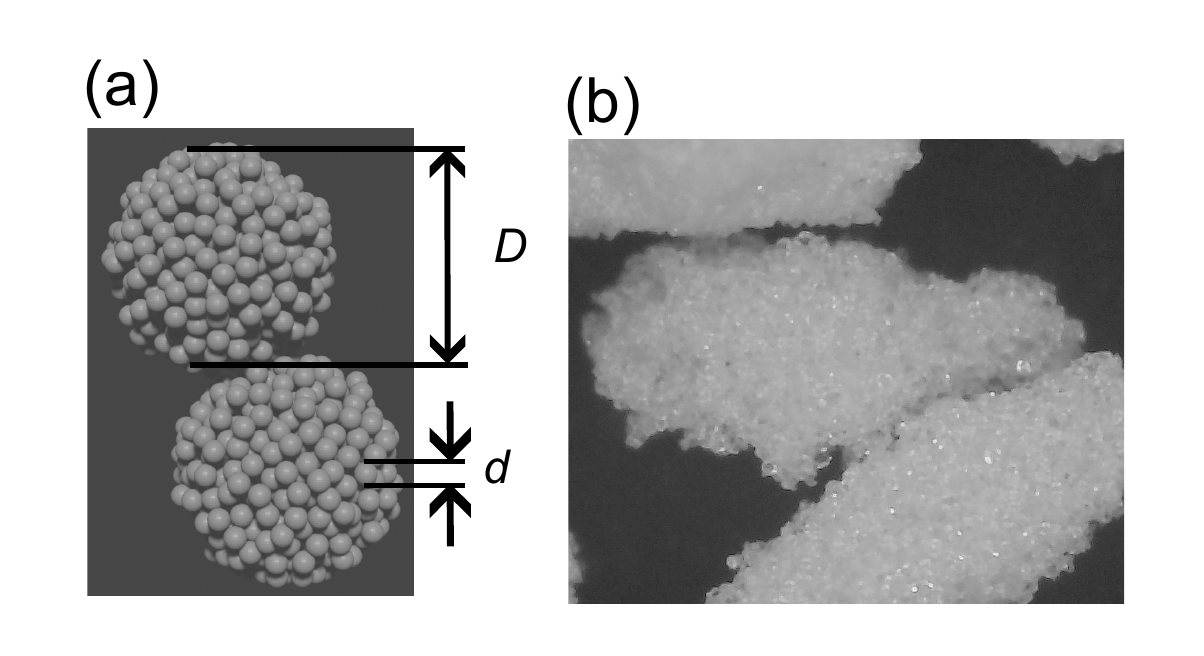}
\caption{(a)Conceptual model of spherical agglomerates with the size of
\(D\). The primary particle size is denoted as \(d\). (b)Photograph of
some agglomerates of hi-d18(S) sample. The width of this picture
corresponds to \(420\mu\mathrm{m}\).}\label{fig:photo}
\end{figure}

The hierarchical and non-hierarchical granular media and the measurement
system used in this study were the same as those used in previous
studies \citep{Yasuda2023}. There are two essential length scales. One
is the diameter of primary particles,\(d\), and the other is the
diameter of secondary particles, \(D\) as illustrated in
Fig.~\ref{fig:photo} (a). Glass beads were used as primary particles.
The non-hierarchical samples, denoted as ``non-hi'', are commercially
available glass beads with typical diameters of 5, 18, and
\(400\mu\mathrm{m}\) (Potters-Ballotini Co., Ltd.: EMB-10, P-001; As One
Corp.: BZ-04). The last sample, non-hi(400), is used only for
porosimetry because its pore size is so large that we can not neglect
the effect of gravity on liquid water transport.

We obtained sintered glass by heating the dried non-hi 18\(\mu\)m
particles in an alumina ceramic container for 50-90 minutes in a muffle
furnace set at \(650^{\circ}{\mathrm C}\). A mortar was used to obtain
agglomerates as the secondary particles by lightly breaking the sintered
glass. We prepared different types of hierarchical granular samples,
denoted as ``hi-d18'', by sieving and named them based on the sieve
used: XS: \(74-250\mu\mathrm{m}\), S: \(250-840\mu\mathrm{m}\), M:
\(840-2000\mu\mathrm{m}\) and L: \(2000-4760\mu\mathrm{m}\). A
photograph of some agglomerates of the hi-d18(S) sample is shown in the
Fig.~\ref{fig:photo} (b) as an example.

Table I shows the increase of the porosity, \(\epsilon\), in hi-d18 from
non-hi samples due to the macroscopic pore. Because we can neglect neck
formation between the connecting beads, the pore structure in the
agglomerates is equivalent to that of the non-hierarchical raw material
before heating. Then, we can calculate microscopic porosity,
\(\epsilon_\mathrm{m}\), as
\begin{equation}\phantomsection\label{eq:non}{
\epsilon_\mathrm{m}:= \epsilon_\mathrm{non}(1-\epsilon)/(1-\epsilon_\mathrm{non}), 
}\end{equation} where \(\epsilon_\mathrm{non}\) is porosity of the
corresponding non-hierarchical granular bed, ``non-hi(18)''. Details of
the sample preparation process are available in previous reports
\citep{Yasuda2023, okubo}.

\begin{figure}
\centering
\includegraphics[width=0.5\textwidth,height=\textheight]{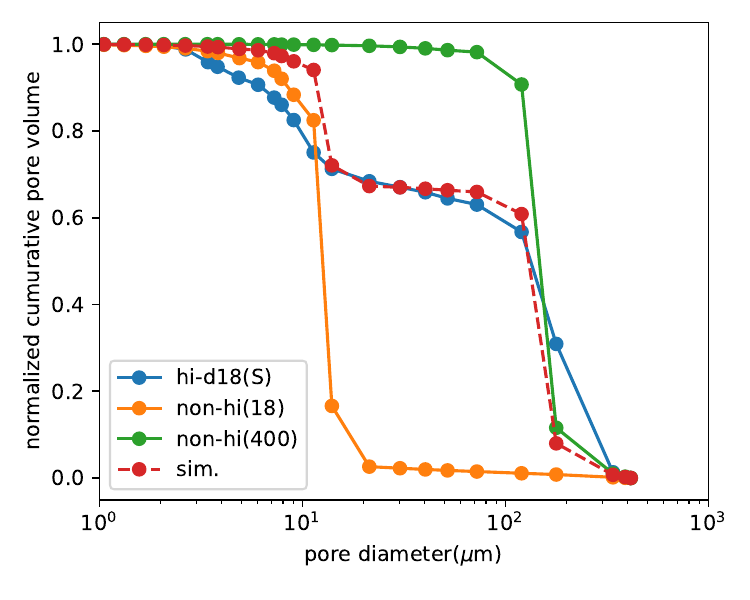}
\caption{The normalized cumulative pore volume of hi-d18(S) and non-hi
samples are plotted together with non-hi samples with diameters of 18
and 400 \(\mu\mathrm{m}\). The red dashed curve denoted as ``sim.''
shows their linear combination.}\label{fig:poresize}
\end{figure}

Fig.~\ref{fig:poresize} shows the normalized cumulative pore volume of
hi-d18(S) together with non-hi samples with diameters of 18 and 400
\(\mu\mathrm{m}\) measured by mercury intrusion porosimetry (MIP). The
non-hi samples for porosimetry were exceptionally sintered in the same
way as the hierarchical samples without being broken so that they could
be measured with MIP. Specifically, The two representative pore
diameters were approximately two to four times smaller than the
constituent particle sizes, \(d\) and \(D\)
\citep{Yasuda2023, Giesche2006}. The larger pores corresponding to \(D\)
are called macroscopic pores and constitute mainly inter-agglomerate
pores. In contrast, the smaller pores corresponding to \(d\) are called
microscopic pores and constitute mainly the pore space within the
agglomerates\citep{Yasuda2023}. The red dashed curve shows a simulation
of the non-hierarchical samples' data, which is a linear combination of
macroscopic and microscopic porosity. The macroscopic pore introduced by
breaking the sintered glass is found to be almost equivalent to
spherical granular bodies with a diameter of 400 µm. The distribution of
microscopic pore size after breaking appears to have changed slightly
compared to before destruction, but in this study, it is considered to
be equivalent.

\subsection{Experimental setup}

We poured 100 g of hi-d18 and non-hi samples into a stainless-steel
sieve mounted in a cylindrical frame with a sheet of filter paper at the
bottom to construct a granular bed. The inner diameter of the container
was 75 mm. 100 g of water was sprayed on the granular bed for 4 min. The
weight of the hydrated granular bed, including the container, and the
weight of the drained water were logged using two electronic balances
(A\&D Co., Ltd.: EK-300i) and a PC with a sampling interval of 5
seconds. These systems, excluding the PC, were maintained in an
incubator (Isuzu Seisakusho Co.Ltd.: VTR-115) at
\(35^{\circ}{\mathrm C}\). This measurement system was the same as that
used in previous studies \citep{Yasuda2023}.

Because the drainage stopped a few seconds after water spraying was
completed, we changed the sampling interval to 1 minute to measure the
weight changes of the granular beds over an extended period due to
drying. The relative humidity in the incubator during drying fluctuated
in the 16-32\% range.

The drying rate can be determined from the time variation of the weight
of the granular beds. Because the circulating air in the incubator hits
the electronic balance, the output value of the electronic balance
exhibits a more significant fluctuation than its specification. We
performed the same runs three times to check the reproducibility of the
data and to reduce the effect of this fluctuation under all experimental
conditions.

\section{Results and Analyses}

\subsection{Drying curves}

\begin{figure}
\centering
\includegraphics[width=0.5\textwidth,height=\textheight]{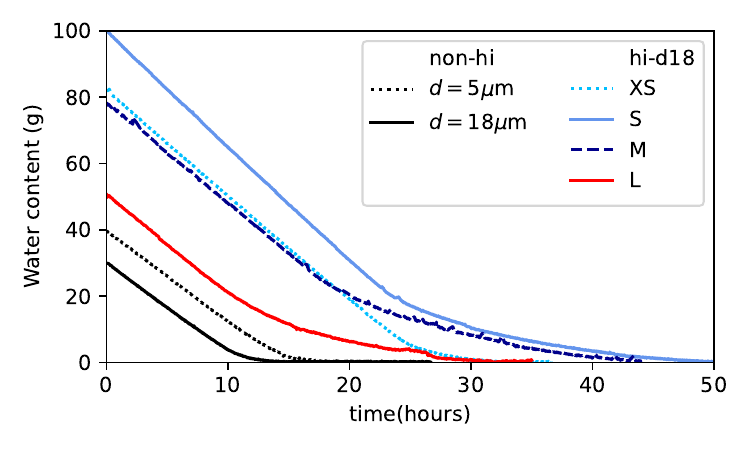}
\caption{Time variation of water content in non-hi and hi-d18 samples
during drying.}\label{fig:rawdata}
\end{figure}

\hyperref[fig:rawdata]{Fig.~\ref{fig:rawdata}} shows an example of the
temporal variation in water content in granular beds due to drying.
Initially, the hierarchical granular beds retained more water than the
non-hierarchical granular beds because of the water retention of the
macroscopic pores \citep{Yasuda2023}. Because most of the water content
decreases at a constant rate that is almost independent of grain size
and structure, it is clear that the most significant factor in the
difference in the time required to dry the granular beds (of the order
of hours) among the samples is the initial amount of water retained.

\subsection{drying rate curves}

We performed least-squares linear fitting to the 30-minute data
(\(t-15\) to \(t+14\) ) to obtain the slope of
\hyperref[fig:rawdata]{Fig.~\ref{fig:rawdata}} with reduced noise.
Because the top and bottom surfaces of the granular bed were open, we
calculated the drying rate defined as the equivalent liquid velocity,
\(V_\mathrm{raw}\), as a function of time, \(t\) by dividing the slope
by \(2\rho a\), where \(a\) is the bed area. The effects of humidity
fluctuations were then corrected. The humidity measured at each time,
\(RH(t)\), was used to calculate the drying rate at a typical humidity
of \(25\%\). A preliminary experiment confirmed that the drying rate
from bulk water is proportional to \(100-RH\) in our incubator.
Therefore, we can calculate the drying rate from our samples, \(V(t)\),
at a typical humidity as follows:
\begin{equation}\phantomsection\label{eq:ERdef}{ 
V(t)= V_\mathrm{raw}(t)\frac{100-25}{100-RH(t)}.
}\end{equation}

We can also calculate the spatially averaged water saturation of the
entire granular bed, \(\langle S\rangle\), from the granular bed packing
fraction. Non-hierarchical granular beds with large values of \(d\)
(2000 and 3000\(\mu\)m) did not exhibit a clear CRP. This trend is
expected because the capillary lengths of these grain sizes roughly
correspond to the sample height. Other non-hierarchical and hierarchical
granular beds show a constant drying rate, at least in the region of
\(\langle S\rangle>0.5\). Therefore, we adopted the average drying rate
of CRPs (\(V_\mathrm{0}\)) as the average drying rate within a range of
0.5\textless{}\(\langle S\rangle\) \textless0.8. For non-hi
(\(d=3000\mu\mathrm{m}\)), there is no data in such a region, and we
assume \(V_\mathrm{0}\) to be that of non-hi(\(d=2000\mu\mathrm{m}\)).
The obtained values of \(V_\mathrm{0}\) were within a narrow range of
0.29-0.35 mm/hour.

\begin{figure}
\centering
\includegraphics[width=0.5\textwidth,height=\textheight]{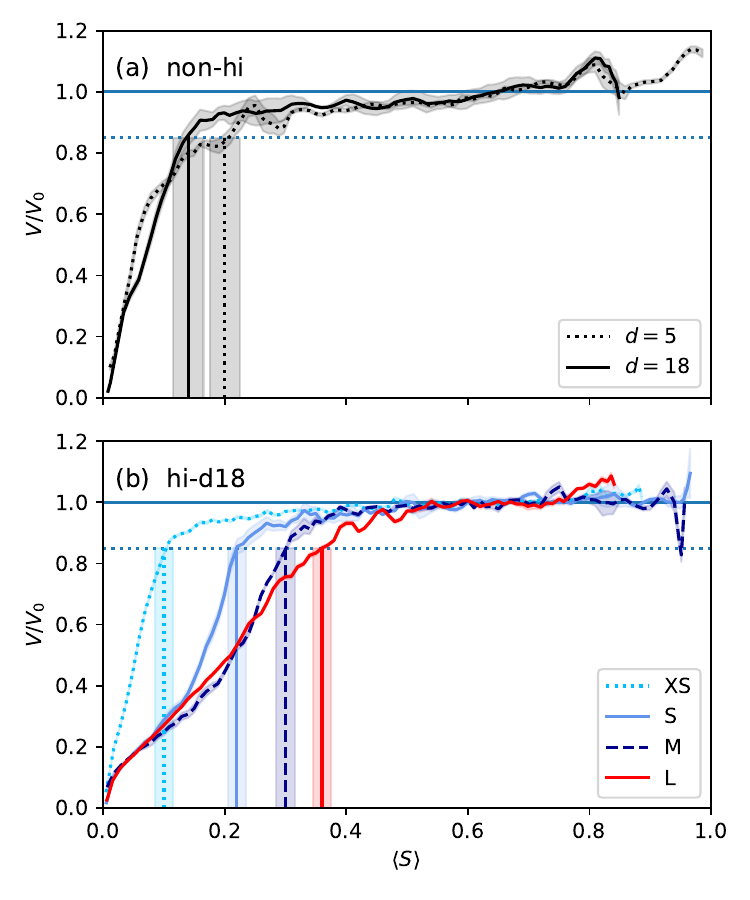}
\caption{drying rate curves normalized by the initial drying
rates,\(V_\mathrm{0}\) for non-hi (a) and hi-d18 (b) as a function of
spatially averaged water saturation, \(\langle S\rangle\). We performed
bin averaging with a constant bin width of 0.05 for non-hi and 0.03 for
all others to obtain these values with less fluctuation. However, the
centers of the bins were swept with a minor step of 0.01 to smooth the
plotted curve. In addition, the vertical dashed lines indicate the
transition points at which the drying rate reaches 0.85 times the
average drying rate of CRP. Thinly painted areas on both sides of each
dashed line indicate the standard deviation of bin
averaging.}\label{fig:ERvsSw}
\end{figure}

The curves in \hyperref[fig:ERvsSw]{Fig.~\ref{fig:ERvsSw}} show the
normalized drying rate \(V/V_\mathrm{0}\) as a function of
\(\langle S\rangle\). To reduce the remnant fluctuation in the drying
rate curves, we performed bin averaging with a constant bin width on the
abscissa (\(\langle S\rangle\) ). We set the bin width to 0.05 for
non-hi and 0.03 for hi-d18. The solid and dashed curves indicate the
average values. The widths above and below the curves represent the
standard deviations. The drying rate fluctuated over the standard
deviation. This fluctuation is not likely due to the noise of the
electronic balance but to some random characteristics of the sample's
evaporation or fluctuations of the air velocity in the incubator.

In addition, FRP was defined as a period when the drying rate is less
than 0.85 times \(V_\mathrm{0}\). The vertical dashed lines in
\hyperref[fig:ERvsSw]{Fig.~\ref{fig:ERvsSw}} indicate transition points.
The thinly painted area on both sides of each dashed line indicates half
of the bin width as the transition point error.

In non-hi, the determined transition points, \(\langle S \rangle^{*}\),
decrease monotonically with increasing \(d\). This trend is similar to
that reported in \citep{Thiery:2017} for granular beds from micrometer
to nanometer scale. It is plausible that the capillary pressure is
inversely proportional to the particle size, whereas water permeability
is proportional to the square of particle size. In coarser granular
materials, the significant permeability continues to lower the water
saturation, facilitates the water supply to the surface, and maintains
CRP.

The transition points of hi-d18 showed a monotonous increase with \(D\).

\subsection{Transition point and drying timescale}\label{sec:section}

We now proceed with further analysis by introducing the volume fraction
of micro-porosity to the total porosity,
\(\gamma^{-1}:=\epsilon_\mathrm{m}/\epsilon\). When the water saturation
is equal to \(\gamma^{-1}\), if the water flow from the macroscopic
pores to the microscopic pores owing to the capillary pressure
difference is sufficiently rapid, the macroscopic pores are considered
almost empty, and the microscopic pores are fully saturated. In
addition, when the water saturation was less than \(\gamma^{-1}\),
\(\gamma\langle S\rangle\) means the degree of water saturation in the
microscopic pores. For non-hierarchical granular beds,
\(\gamma^{-1}=1\).

\begin{figure}
\centering
\includegraphics[width=0.5\textwidth,height=\textheight]{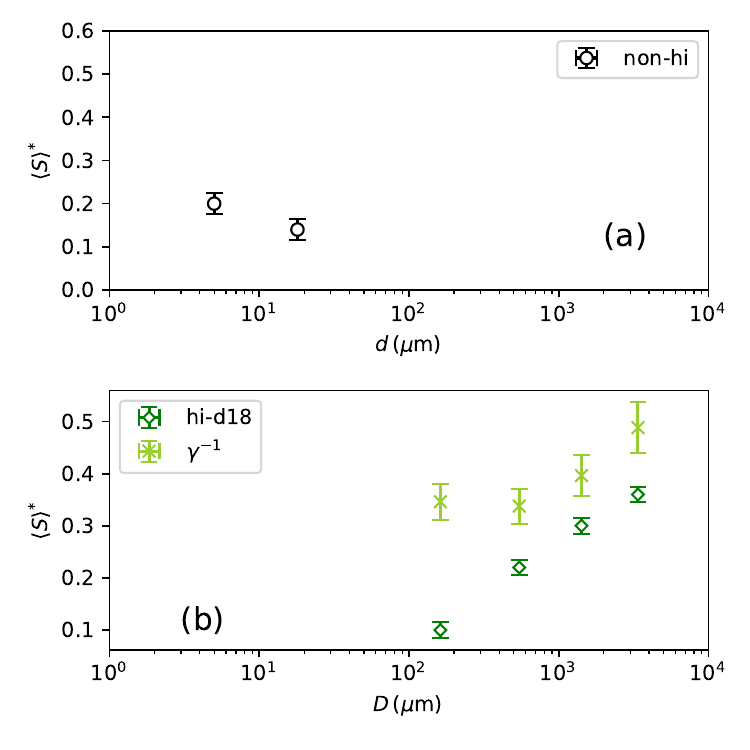}
\caption{Transition points (\(\langle S\rangle^{*}\)) and particle sizes
(\(d\) and \(D\)) for hierarchical and non-hierarchical granular beds.
Panel (a) shows non-hi with the x-axis of the primary particle size,
\(d\). Panel (b) shows hi-d18 with the x-axis of the secondary particle
size, \(D\). For comparison, \(\gamma^{-1}\) are
shown.}\label{fig:SwvsdD}
\end{figure}

The calculated \(\gamma^{-1}\) values are shown in
\hyperref[fig:SwvsdD]{Fig.~\ref{fig:SwvsdD}}. We adopt the mid-range of
the agglomerate size as the representative value, \(D\), of hi-d18. For
all the hierarchical granular beds, the transition points,
\(\langle S \rangle^{*}\), were smaller than \(\gamma^{-1}\). This means
that when \(\langle S\rangle\) reaches \(\gamma^{-1}\), the water flow
to the bed surface is sufficient to maintain CRP. That is, the water in
the microscopic pores is not completely isolated by the air-filled
macroscopic pores.

\begin{figure}
\centering
\includegraphics[width=0.5\textwidth,height=\textheight]{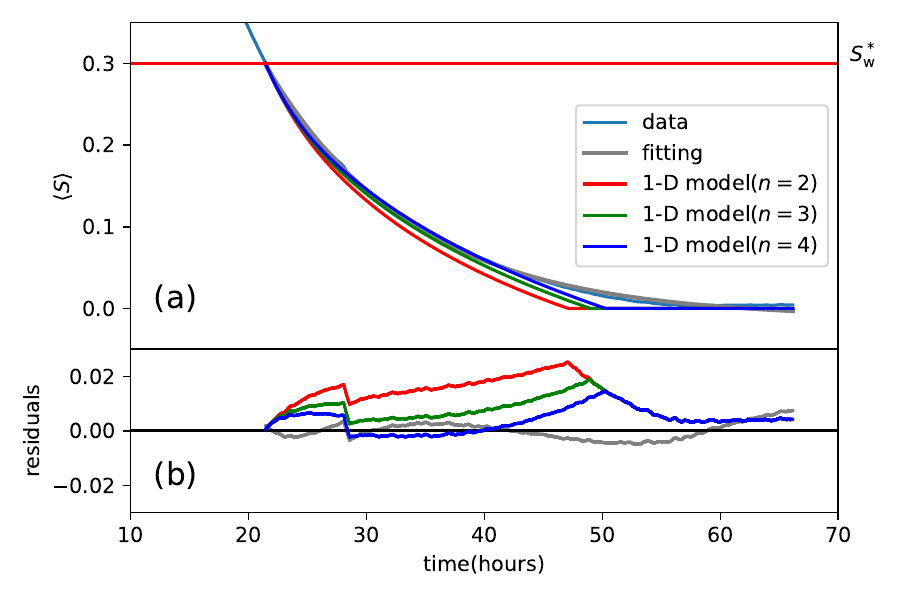}
\caption{Time validation curve of the second run of M-sized sample of
hi-d18 as an example of drying time constant, \(\tau\), determination.
(a): The gray curve is optimized with the function:
\(\langle S\rangle=A\times\exp(-t/\tau)+C\) to fit the experimental
data, a blue solid curve. Dashed curves are theoretically calculated
with Eq.~\ref{eq:kinjikai3} and Eq.~\ref{eq:kinjikai4}. (b): residue for
the optimization and theoretical curves.}\label{fig:analysis}
\end{figure}

We can define the saturation time, \(t^{*}\), when saturation reaches
the transition point, \(\langle S \rangle^{*}\).
\hyperref[fig:analysis]{Fig.~\ref{fig:analysis}}(a) shows one of the
drying curves during the FRP, \(t>t^{*}\), of M-sized samples with
\(d\)=18\(\mu\)m theoretical curves and the fitting curve of an
exponential function:
\begin{equation}\phantomsection\label{eq:exponential}{
\langle S\rangle = A\times \exp(-t/\tau) + C, 
}\end{equation} where \(A\), \(C\), and \(\tau\) are the fitting
parameters. The theoretical curves are calculated with
Eq.~\ref{eq:kinjikai2} and Eq.~\ref{eq:V}. The residue of the fitting
curve is not like random noise but shows a systematic change (b).
However, it is smaller than the residue of any of the theoretical
curves. The change in saturation appears to be approximated by an
exponential function better than our model. The obtained values of
\(\tau\) are a measure of the drying time in the FRP, which we call the
drying time constant.

\begin{figure}
\centering
\includegraphics[width=0.5\textwidth,height=\textheight]{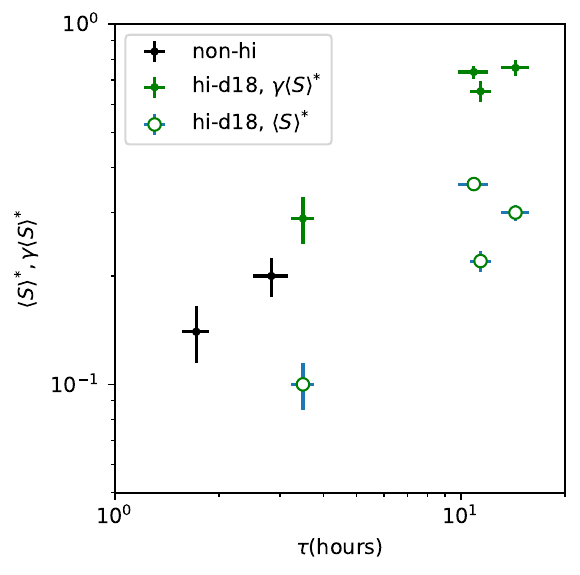}
\caption{Total and Microscopic pore saturation at the
transition,\(\langle S \rangle^{*}\) and
\(\gamma \langle S \rangle^{*}\) are plotted versus the drying time
constant, \(\tau\).}\label{fig:Swvstau2}
\end{figure}

\hyperref[fig:Swvstau2]{Fig.~\ref{fig:Swvstau2}} shows the correlation
of the drying time constant,\(\tau\), to the total and the microscopic
pore saturation at the transition, \(\langle S\rangle^{*}\) and
\(\gamma \langle S\rangle^{*}\). The transition point,
\(\langle S \rangle^{*}\), of hierarchical granular bodies is on a
different trend from non-hierarchical granular bodies, while the
spatially averaged micro-saturation at the transition,
\(\gamma \langle S \rangle^{*}\), appears to be on the same trend as the
transition point of non-hierarchical granular bodies. This result
suggests that for hierarchical granular bodies, microscopic saturation
is essential for the transition to FRP and for the drying rate in FRP
than saturation.

\section{Discussions}

\subsection{drying rate}

\begin{figure}
\centering
\includegraphics[width=1\textwidth,height=\textheight]{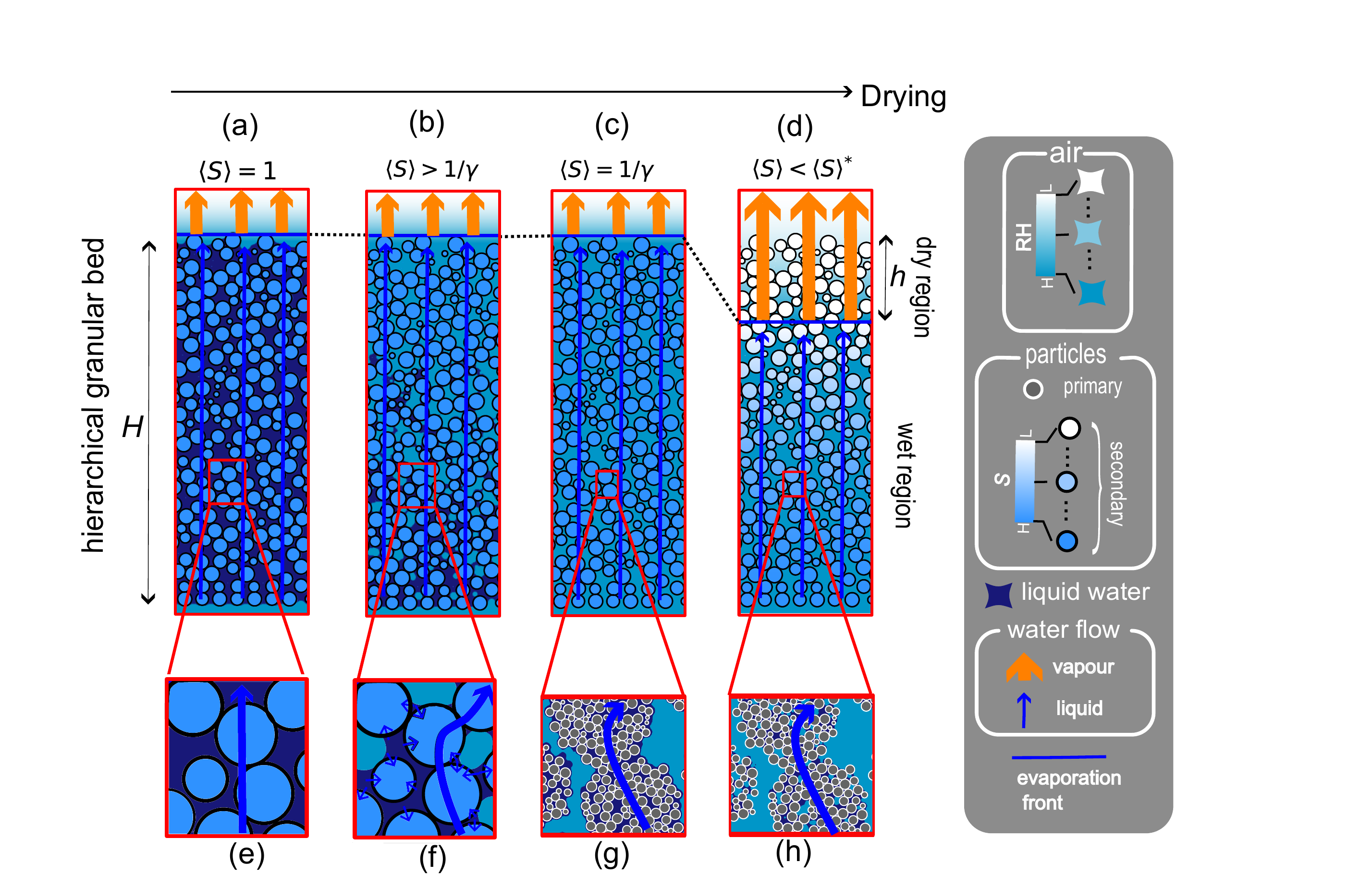}
\caption{Conceptual diagram of the drying process of a hierarchical
granular bed as cross-section view.(a)-(c), (e)-(g) represent CRP, while
(d) and (h) represent FRP. At first, both macroscopic and microscopic
pores are filled with water, (a) and (e). At this stage, liquid water
flow can path through both pores. At \(\langle S \rangle > 1/\gamma\),
macroscopic pore water between secondary particles remains, (b), but is
being absorbed into the microscopic pore by the capillary pressure
gradient created by the upward liquid water flow as shown in (f). At
\(\langle S \rangle = 1/\gamma\), all liquid water is present in the
microscopic pore, which is saturated by water. From this point onwards,
liquid water flow is limited in the microscopic pore and the microscopic
pore becomes unsaturated and less permeable as shown in (g) and (h). In
FRP, the microscopic pore in secondary particles in the dry region are
dried and do not provide a pathway for liquid water, but do provide a
pathway for water vapour diffusing from the evaporation
front.}\label{fig:gainen2}
\end{figure}

As shown in \hyperref[fig:ERvsSw]{Fig.~\ref{fig:ERvsSw}}, we observed
CRP and FRP in almost all hierarchical granular beds. Considering the
typical humidity in the chamber of 25\%, the initial drying rate during
CRP, \(V_\mathrm{0}\), is given as\citep{Thiery:2017}:

\begin{equation}\phantomsection\label{eq:vapor}{
V_\mathrm{0} = \frac{0.75 \rho_{0}D_\mathrm{V} }{\rho\delta},
}\end{equation} where \(\rho_{0}=40\mathrm{g/m}^{3}\) is the density of
water vapor at the temperature of \(35^{\circ}\mathrm{C}\), and
\(D_\mathrm{V}=2.7\times10^{-5}\mathrm{m^{2}/s}\) is the diffusion
coefficient of water vapor at \(35^{\circ}\mathrm{C}\). Based on the
experimental values of \(V_\mathrm{0}\), the values of \(\delta\) were
calculated within a range of 8.3 - 9.9 mm. The narrow range indicates
slight variations in air velocity in the incubator during the
measurement. In addition, we observed no significant decrease in the
drying rate in the saturation range of
\(\langle S\rangle > \gamma^{-1}\). This implies that the drying rate in
CRP can be maintained only by water vapor diffusion from microscopic
pores exposed to the granular bed surfaces. This idea is consistent with
the fact that the estimated macroscopic pore size, a measure of the size
of blank space of the vapor supplying microscopic pore, is about
\(D/4 - D/2\) which is smaller than the boundary layer thickness,
\(\delta\). In Fig.~\ref{fig:gainen2}, we illustrated the distribution
and flow of water at each step based on our experimental results.

\subsection{Power law of the transition point}

\begin{figure}
\centering
\includegraphics[width=0.5\textwidth,height=\textheight]{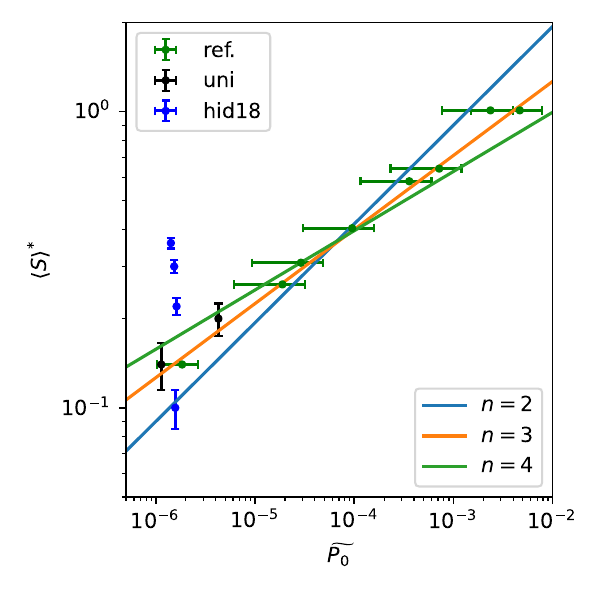}
\caption{Log-log plot with \(\widetilde{P_{0}}\) on the horizontal axis
and \(\langle S \rangle^{*}\) on the vertical axis. The green dots
(ref.) are data read from a figure in the reference \citep{Thiery:2017}.
However, \(\langle S \rangle^{*}\) was read directly from the drawing.
Since our sample (uni and hi-d18) evaporated from both the top and the
bottom sides, half of the sample height was assigned to \(H\). Also, as
the values of \(H\) and \(V_{0}\) are only given as ranges in the
ref.\citep{Thiery:2017}, we used error bars to represent their effects.
The three straight lines (blue, orange, and green) correspond to assumed
\(n\) values (2,3,4); the three lines (blue, orange, and green)
correspond to the assumed \(n\) values (2,3,4). However, the
proportionality constants were determined to pass near the data points
other than hi-d18.}\label{fig:power_law}
\end{figure}

The relationship between \(\widetilde{P_{0}}\) and
\(\langle S \rangle^{*}\) is shown in Fig.~\ref{fig:power_law} based on
our data and those of Thiery et al.~Since evaporation occurred from both
the top and bottom sides of our sample, half of the sample height was
substituted for \(H\). As the values of \(H\) and \(V_{0}\) are not
given for each sample in the report \citep{Thiery:2017}, the effect of
the distribution of the given values is expressed as an error bar.

Although \(\widetilde{D_\mathrm{L}^{0}}\) still depends on the pore
structure, it has a narrow range (\(0.42-0.51\)) of porosity
\(\epsilon\) for non-hierarchical samples, so we assume
\(\widetilde{D_\mathrm{L}^{0}}\) to be a constant for non-hi and
``ref.'' Three lines in Fig.~\ref{fig:power_law} show the model equation
\(\langle S \rangle^{*} \propto \widetilde{P_{0}}^{\frac{1}{n+1}}\)
assuming 2, 3, and 4 as candidate values of \(n\) passing near the data
points except for data of hi-d18. The data points other than hi-d18 lie
approximately on the line corresponding to \(n=3\). Since this value is
also consistent with Eq.~\ref{eq:thiery}, ``n'' is assumed to be three
thereafter in this paper. From the slope, the following empirical values
are obtained. \begin{equation}\phantomsection\label{eq:4point4}{
\widetilde{D_\mathrm{L}^{0}}^{-\frac{1}{4}}\sigma^{*}= 4.0
}\end{equation} In order to examine the validity of these values,
numerical solutions for Eq.~\ref{eq:model1} and Eq.~\ref{eq:model2} were
obtained (Appendix A). We can obtain \(\sigma^{*}= 0.69\). In this
appendix, it is also confirmed that Eq.~\ref{eq:kinjikai2} and
Eq.~\ref{eq:kinjikai3} are good enough approximations for the
calculations. Appendix B also provides estimates of
\(\widetilde{D_\mathrm{L}^{0}}\) calculated from existing data.
Substituting these into the left-hand side of Eq.~\ref{eq:4point4}
yields a value of 2.4. This value is not fully consistent with the value
on the right-hand side of Eq.~\ref{eq:4point4}, suggesting that our
model is not sufficiently precise to explain the experimental results
quantitatively, and further improvements are needed. Substitution of the
value into Eq.~\ref{eq:kinjikai} and Eq.~\ref{eq:sol1}, we can see that
\(m\), a constant value defined in Eq.~\ref{eq:critical}, is 6.6 and
confirm that it is consistent with its definition, ``not sufficiently
smaller than 1''.

The larger values of the transition point for the hi-d18 samples mean
small values of \(\widetilde{D_\mathrm{L}^{0}}\), which may be due to a
reduction in their permeability due to the dried macroscopic pore.

\subsection{The drying rate in the FRP}

As seen in Fig.~\ref{fig:analysis}, Eq.~\ref{eq:kinjikai3} and
Eq.~\ref{eq:kinjikai4} do not fully reproduce the drying process in FRP.
This is presumably due to the fact that the spatial uniformity of
\(D_\mathrm{L}\), which is assumed as a precondition for these
equations, is not fully established. Before adding inhomogeneity to the
model, let us see how far Fig.~\ref{fig:ERvsSw} can be reproduced with a
uniform model.

\begin{figure}
\centering
\includegraphics[width=0.5\textwidth,height=\textheight]{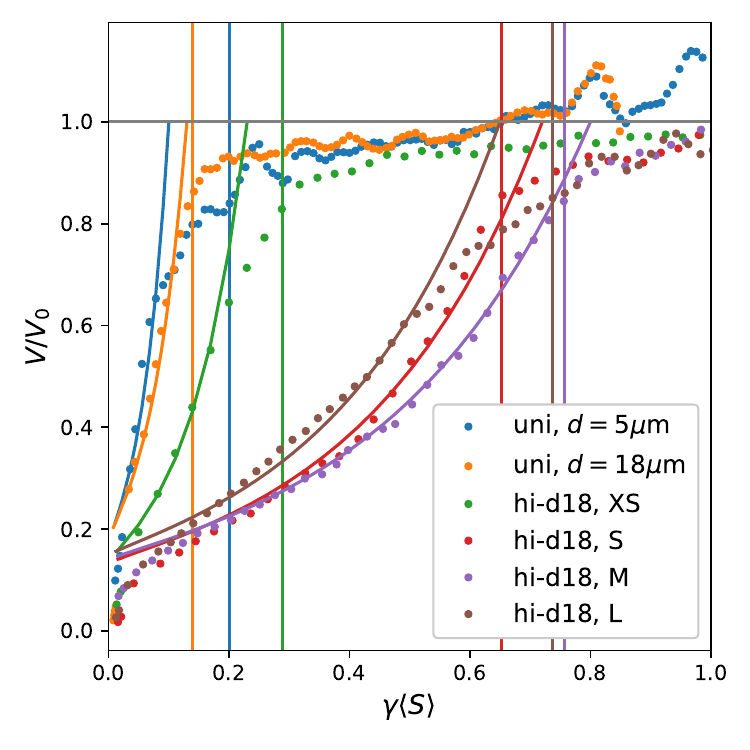}
\caption{The dotted line represents the falling drying rate at FRP of
our sample (uni, hi-d18). The horizontal axis is the microscopic
saturation. The vertical solid lines show the transition
(\(\gamma \langle S \rangle^{*}\)) to FRP in the corresponding color
data. The curves show the theoretical values with their transition point
set to fit the corresponding color data during FRP.}\label{fig:FRP_ER}
\end{figure}

In Fig.~\ref{fig:FRP_ER}, the experimental data for the saturation
dependence of the drying rate were compared with a theoretical curve
with \(n=3\). However, when calculating the theoretical curve, the
transition points were set to follow the downward convex behavior of the
experimental data at low saturation instead of adopting the experimental
data. Both the theoretical and experimental curves are characterized by
a downward convex curve that asymptotically approaches \(1/(1+N)\) at
low saturation. In addition, two main differences can be observed.
First, at very low saturations below 0.1, only the experimental values
show a reduction in the drying rate beyond the asymptote. This may be
due to the fact that at these very low saturations, most of the water is
present in isolated bridges between the particles, so the assumption
that one-dimensional water vapor diffusion is the rate-limiting factor
Eq.~\ref{eq:V} does not hold. Secondly, compared to the theoretical
curve, the experimental data do not show a rapid decrease in the drying
rate around the transition point. Such a degradation of the curve is
common in many types of spectroscopy and is known as inhomogeneous
spectral broadening, so our case is analogous to a change originating
from inhomogeneity. The bin averaging process used to obtain the
experimental values should have resulted in a bin-width-induced
degradation, which should also be responsible for both differences.

\begin{figure}
\centering
\includegraphics[width=0.5\textwidth,height=\textheight]{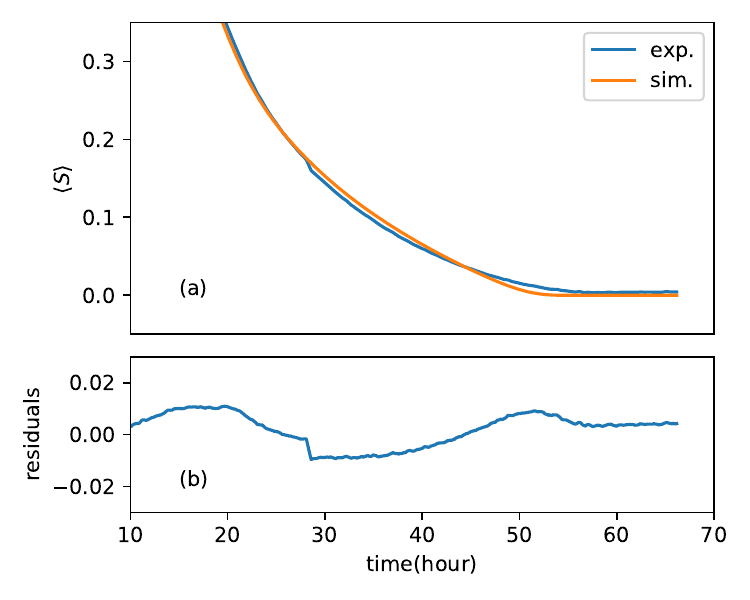}
\caption{(a): Time variation of water saturation in the M-sized sample
of hi-d18. We can better fit the simulated curve, ``sim.'', to the
experimental data, ``exp.'', by introducing an inhomogeneous transition
point. The assumed distribution of transition points is shown in
Fig.~\ref{fig:FRP_ER3} as a bar chart. (b): The residuals mean the
difference between ``sim.'' and ``exp.''}\label{fig:FRP_ER2}
\end{figure}

\begin{figure}
\centering
\includegraphics[width=0.5\textwidth,height=\textheight]{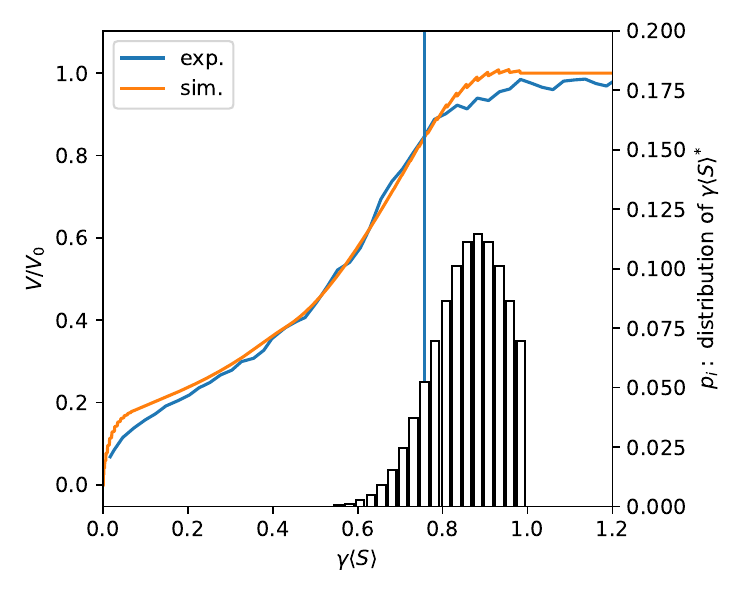}
\caption{The dependence of the drying rate on water saturation in the
M-sized sample of hi-d18. We can better fit the simulated curve,
``sim.'', to the experimental data, ``exp.'', by introducing an
inhomogeneous transition point. The assumed distribution of transition
points is shown as a bar chart.}\label{fig:FRP_ER3}
\end{figure}

To qualitatively assess the effects of inhomogeneity, horizontal
inhomogeneity of the transition point is introduced. In this case,
saturation is no longer a one-dimensional distribution. For simplicity,
horizontal advection and diffusion are ignored. In this case, the time
variation of the water saturation \(\langle S\rangle\) can be written by
a linear combination of the theoretical curves presented in
Fig.~\ref{fig:analysis} as follows:
\begin{equation}\phantomsection\label{eq:hetero}{
\langle S\rangle(t)=\sum_{i=0}^{i_\mathrm{max}} p_{i}\overline{S_{\mathrm{w}, i}}(t), 
}\end{equation} where \(p_{i}\) is the probability that the transition
point takes \(S_{\mathrm{w}}^{*}\) and \(\overline{S_{\mathrm{w}, i}}\)
is the time variation of water saturation in that case.
Fig.~\ref{fig:FRP_ER2} and Fig.~\ref{fig:FRP_ER3} show the results of
calculation in which the normal distribution is introduced as the first
trial function for inhomogeneous distribution as an example. The
experimental data of interest are the M-sized sample of hi-d18 used in
Fig.~\ref{fig:analysis}. The assumed distribution of transition points
is shown in Fig.~\ref{fig:FRP_ER3} as a bar chart. Here,
\(\gamma \langle S \rangle^{*}<1\) was assumed in order to maintain the
condition that the macroscopic pore is dry at the transition point. It
can be confirmed that the time variation of saturation,
Fig.~\ref{fig:FRP_ER2}, and the saturation dependence of the drying rate
Fig.~\ref{fig:FRP_ER3} can be brought closer to the experimental results
by introducing heterogeneity at the transition point. It was also found
that the introduced inhomogeneity can also explain the lower drying rate
beyond the asymptote in the low saturation region. Under the assumption
of horizontal heterogeneity, areas with small transition points dry out
completely earlier than areas with large transition points. In other
words, the reduction of the drying rate beyond the asymptote may be
attributed to the fact that the evaporating area becomes smaller at low
saturation.

\subsection{Drying time constant}

With the definition of \(\gamma\), Eq.~\ref{eq:prop} can be rewritten as
\begin{equation}\phantomsection\label{eq:prop2}{
\tau \frac{V_{0}}{\epsilon_{m} H}: \gamma \langle S \rangle^{*}=
\tau \frac{V_{0}}{\epsilon H}: \langle S \rangle^{*} = const.
}\end{equation}

\begin{figure}
\centering
\includegraphics[width=0.5\textwidth,height=\textheight]{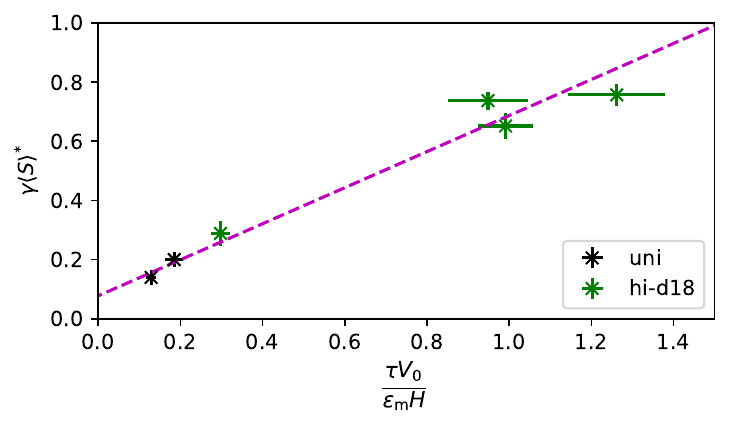}
\caption{Non-dimensionalized drying time
constant\(\frac{\tau V_{0}}{\epsilon_m H}\) and microscopic pore
saturation at transition \(\gamma\langle S\rangle^{*}\). The straight
line is the fitting line obtained by the weighted least squares
method.}\label{fig:Swvstau3}
\end{figure}

\hyperref[fig:Swvstau3]{Fig.~\ref{fig:Swvstau3}} shows the linear
relationship described in Eq.~\ref{eq:prop2} with a small intercept in
the axis of \(\gamma S\) at less than 0.1. The saturation degree
corresponds to the residual water, \(S_\mathrm{r}\), measured for many
granular materials are in this region, so the slight discrepancy with
this model may be due to the fact that the model ignores
\(S_\mathrm{r}\). We therefore propose the following empirical formula
from the approximate straight line in
\hyperref[fig:Swvstau3]{Fig.~\ref{fig:Swvstau3}}.
\begin{equation}\phantomsection\label{eq:keiken}{
\gamma \langle S \rangle^{*}=0.61\times \tau \frac{V_{0}}{\epsilon_{m} H}+0.08,
}\end{equation} where the two constants (0.61 and 0.08) would depend on
the structure of the micro-pore, the former being mainly influenced by
the heterogeneity of the pore structure and the latter by
\(S_\mathrm{r}\).

\section{Conclusion}

The proposed model, as presented in Eq.~\ref{eq:model1} and
Eq.~\ref{eq:model2}, assuming \(n=3\), is not only consistent with the
empirical formula for the transition points of Thiery et
al.\citep{Thiery:2017} Appendix A examines the difference between
approximate and numerical solutions. It is negligible with respect to
the error levels observed in the present experiments.

Based on the model, we could introduce a new parameter,
\(\widetilde{P_0}\), to successfully align transition points of their
data and our data for non-hierarchical samples in a single line in
log-log plots within a margin of error. Although we can estimate the
proportionality constant from existing data on capillary pressure and
water permeability with the same order of magnitude as that of our model
in Appendix B, the difference is not negligible.

On the other hand, in a middle range of FRP, experimental data of the
drying rate show similar behavior to the theoretical curve approaching
the minimum value of the model, \(1/(1+N)\). Differences between the
model and the experiment are evident near the transition point and in
the region of low micro-pore saturation below 0.1, but it is clear that
the discrepancies can be reduced for the case of hierarchical samples by
assuming a simple case where the transition point has horizontal
inhomogeneity.

The correlation between the experimental drying time constant and
micro-pore saturation was obtained by making the drying time constant
dimensionless based on our model, and a universal empirical equation
(Eq.~\ref{eq:keiken}) was obtained.

The validation of the model carried out in this study is not yet
sufficient. Direct proof of the water transport model in hierarchical
and non-hierarchical hydrophilic granular media requires the expansion
of the dry region to be observed with MRI (\citep{Thiery:2017};
\citep{Maillet:2022}) and dyes
(\citep{Shokri2009, Shokri2010, Shokri2011}), together with
quantification of the effect of reduced water permeability in the
hierarchical bed. The differences in the transition point may also be
caused by sample inhomogeneities. However, there are two issues that
remain to be addressed in order to deepen the discussion on the
transition point. The first is the method of defining the transition
point. In this study, the transition point was defined as the point at
which the drying rate was 0.85 times the CRP. This definition was
adopted to reduce the effects of measurement errors, but it is
arbitrary. In order to redefine the transition based on the sample
inhomogeneity, experiments must be conducted to quantitatively clarify
the effect of heterogeneity. In this study, the hierarchical granular
material that suggested heterogeneity has no method for predicting the
rate of decrease in water permeability due to macroscopic pores, so at
present, \(\widetilde{D_\mathrm{L}^{0}}\) is unknown. In addition,
although the value of \(\widetilde{D_\mathrm{L}^{0}}\) was estimated for
our non-hi samples, the effect of \(S_\mathrm{r}\) cannot be ignored,
and it is thought that the application of our model is close to the
limit. Therefore, we propose that the next appropriate step is to
measure the drying rate of non-hierarchical samples with large
\(\widetilde{P_{0}}\) as in the work of Thiery et al., but with further
improved accuracy.

As for the dependence of \(P_\mathrm{c}\) on \(S\), a linear function
was used to approximate the van Genuchten curve in the intermediate
region. It is known that similar curves can be reproduced by numerical
calculations of meniscus size in partially saturated spherical granular
\citep{Sweijen:2017}. Therefore, the consistency of our model with data
from nano-porous samples \citep{Thiery:2017} suggests that the capillary
pressure can be determined from the meniscus size in partially saturated
pores without considering water films on hydrophilic surfaces of a few
nanometers thickness \citep{Nishiyama2021}. On the other hand, it has
been reported that in monolithic nano-porous silica glasses, no dry
regions are observed during the drying process, and the pore water is
present in a water-film-like confined state in the low saturation region
\citep{Maillet:2022}. As this difference may be attributed to
differences in pore geometry, caution should be exercised when extending
our arguments to other nano-porous media.

\section{Declaration of Competing Interest}

The authors declare that they have no known competing financial
interests or personal relationships that could have appeared to
influence the work. The authors declare that they have no known
competing financial interests or personal relationships that could have
appeared to influence the work.

\section{Acknowledgment}

This work was supported by JSPS KAKENHI, Grant No.~JP18H03679,
JP23H04134, and JP24H00196.

\section{Appendix A: Numerical solution of our model}

To obtain the numerical solution, we apply spatial discretization to the
differential equations Eq.~\ref{eq:model1} and Eq.~\ref{eq:model2}. The
thickness of the granular bed (\(0<z' \le 1\)) is divided into
\(M=500\). The saturation values, \(\sigma_{i}(t)\), of the i-th regions
and the flux between the i-th and the neighboring regions, \(u_{i}\),
satisfy the following equations:
\begin{equation}\phantomsection\label{eq:risan1}{
\begin{aligned}
u_{i} &=
\begin{cases}
0 & (i \ge k+1 \text{ or } i=0),\\
\dfrac{1}{1 + N\tilde{h}} & (i = k \text{}),\\
-\left(\frac{\sigma_{i}+\sigma_{i+1}}{2}\right)^{3}
\left[ (\sigma_{i+1}-\sigma_{i})M \right] &
(0 < i < k) ,\\
\end{cases}\\
\tilde{h} &=(M-k)/M,\\
\frac{d \sigma_{i}}{dt'}&=
\begin{cases}
-(u_{i+1}-u_{i})M & (0\le i< k),\\
0 & (i\ge k),
\end{cases}\\
\sigma_{i}&=0 \hspace{1cm} (i\ge k).
\end{aligned}
}\end{equation}

We started the numerical simulation with the initial condition of
\(\sigma_{i}=1\) for \(i=0, ..., M+1\) and \(k=M\). The typical value of
our samples, 6, was used as \(N\). We used the function ``solve\_ivp''
in an open-source library, ``scipy'' \citep{2020SciPy-NMeth}, which
integrates ordinary differential equations with an initial value. We
programmed an event to generate when the top cell of the wet region,
\(\sigma_{k}\), dries out, i.e., reaches the value of \(0\). If the
event occurs, the program updates the boundary conditions by decreasing
\(k\).

\begin{figure}
\centering
\includegraphics[width=0.5\textwidth,height=\textheight]{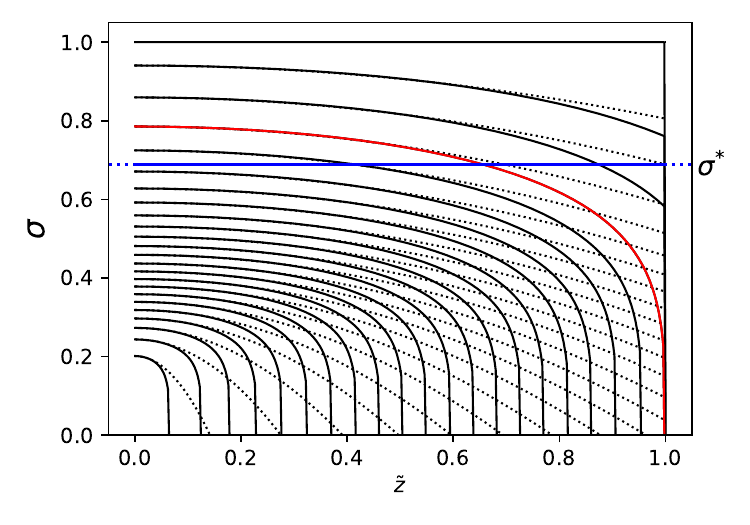}
\caption{Black curves show the time variation of the vertical
distribution of \(s\). The time intervals between neighboring lines were
fixed to be the same value, \(0.1\tau_{0}\). The red curve indicates the
distribution at the transition from CRP to FRP. The blue line indicates
the transition point, \(\sigma^{*}=0.689\), defined as the spatial
average of \(\sigma\) at the transition. Dotted curves are parabolas
fitting to the vertical distributions around the bottom
(\(\tilde{z}\sim 0\)).}\label{fig:distri}
\end{figure}

\hyperref[fig:distri]{Fig.~\ref{fig:distri}} shows snapshots of
\(\sigma_{i}\) distribution with an interval of
\(\Delta \tilde{t}=0.1\). The formation of dry regions is seen beneath
the top surface. We averaged \(\sigma_{i}\) at the transition to get the
following value.
\begin{equation}\phantomsection\label{eq:integral_value}{
\sigma^{*}=0.69.
}\end{equation} Also, we can confirm the spatial distribution of water
in the wet region is close to a parabola as predicted,
Eq.~\ref{eq:parabola}, except for just beneath the evaporating boundary
to the dry region.

\begin{figure}
\centering
\includegraphics[width=0.5\textwidth,height=\textheight]{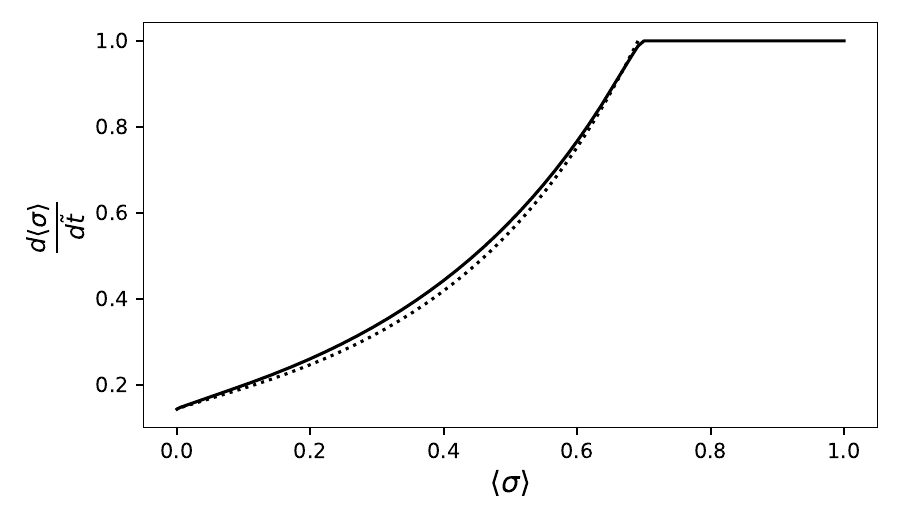}
\caption{The solid curve shows the relationship between
\(\langle \sigma \rangle\) and \(d\langle \sigma \rangle/d\tilde{t}\) .
The dotted curve is a similar plot to the solid curve but is obtained
from Eq.~\ref{eq:kinjikai3} and
Eq.~\ref{eq:kinjikai4}.}\label{fig:numeri}
\end{figure}

From \hyperref[fig:numeri]{Fig.~\ref{fig:numeri}}, we can confirm that
Eq.~\ref{eq:kinjikai3} and Eq.~\ref{eq:kinjikai4} give almost correct
solutions of \(\langle \sigma \rangle\) without solving
Eq.~\ref{eq:model1} and Eq.~\ref{eq:model2} directly.

\section{Appendix B: Evaluation of the non-dimensional diffusion
constant in non-hierarchical samples}

\begin{figure}
\centering
\includegraphics[width=0.5\textwidth,height=\textheight]{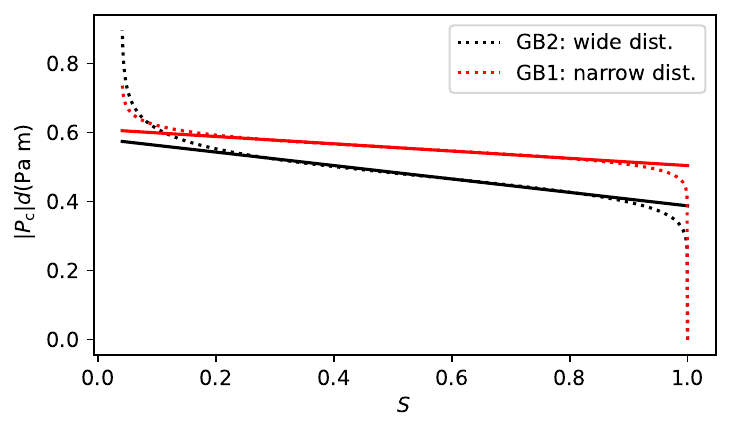}
\caption{The two dotted lines are the van Genuchten curves fitted to
reported experimental values \citep{Culligan:2004, HILPERT2001} of
capillary pressure at various water saturations for non-hierarchical
granular materials (GB2, GB1) consisting of two types of spherical glass
beads \citep{Sweijen:2017}. The solid lines are straight lines that fit
the curves at the range of \(0.2<S <0.8\). The reciprocals of the slopes
are -5.1 (GB2) and -9.5 (GB1), respectively.}\label{fig:Sw_Pc}
\end{figure}

There have been many studies on the capillary pressure,
\(P_\mathrm{c}\)\citep{Sweijen:2017}.
\hyperref[fig:Sw_Pc]{Fig.~\ref{fig:Sw_Pc}} shows the Van Genuchten
curves that approximate the experimental results of the dependence of
capillary pressure on water saturation for two types of glass beads
(GB1\citep{HILPERT2001} and GB2\citep{Culligan:2004}). The capillary
pressure increases rapidly as the water saturation approaches a specific
value, \(S_\mathrm{r}\), and decreases rapidly as the water saturation
approaches 1. Here, \(S_\mathrm{r}\) corresponds to the water in the
isolated bridges, called irreversible water or residual water, and has a
hysteresis value of less than 0.2. This figure shows that the capillary
pressure can be approximated with a straight line in an intermediate
range. GB1 and GB2 had different particle size dispersion widths; the
former had a value of 10\% of the mean diameter, and the latter had a
value of approximately 30\%. The product of pore pressure and average
particle size depends on the dispersion width \citep{Sweijen:2017}.
Since the glass beads used in this experiment had a wide dispersion of
about 50\%, we regard the value of GB2 as suitable for our data.
Similarly, we regard the value of GB1 as suitable for the narrow
dispersion samples used in \citep{Thiery:2017}. Because capillary
pressure should be proportional to the surface energy of the water, we
can describe the dependence of the capillary pressure on water
saturation as follows: \begin{equation}\phantomsection\label{eq:Pc}{
\frac{d P_\mathrm {c}}{dS} = - \frac{\Gamma}{\alpha d},
}\end{equation} where \(\alpha\) is a dimensionless quantity. We can
obtain the value for our samples from the slope of the line (GB2) in
Fig.~\ref{fig:Sw_Pc} as \begin{equation}{
\alpha \sim 0.36.
}\end{equation}

\begin{figure}
\centering
\includegraphics[width=0.5\textwidth,height=\textheight]{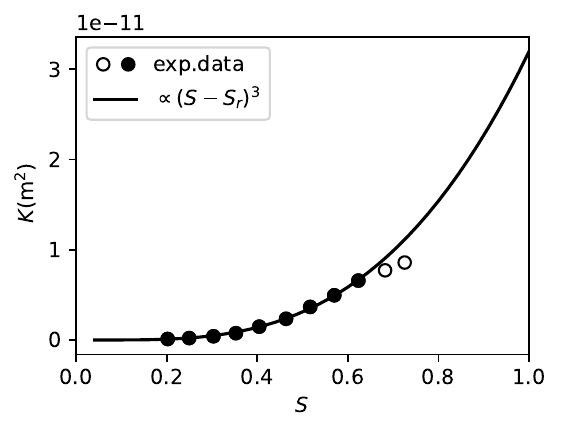}
\caption{The two types of circles indicate experimental water
permeability from non-hierarchical granular materials consisting of
spherical glass beads with an average diameter of
\(180\mu\mathrm{m}\)\citep{agg1}, provided that water viscosity was
assumed to be \(0.001\mathrm{Pa}\cdot\mathrm{s}\). The solid line is the
cubic function, \(\propto (S -S_\mathrm{r})^3\) fitted to the selected
data indicated by the filled circles. We can obtain an optimal value of
0.07 for \(S_\mathrm{r}\) with the least-squares
method.}\label{fig:Sw_K}
\end{figure}

As shown in \hyperref[fig:Sw_K]{Fig.~\ref{fig:Sw_K}}, the water
permeability, K, of non-hierarchical beds reaches zero when the water
saturation approaches \(S_{r}\).
\hyperref[fig:Sw_K]{Fig.~\ref{fig:Sw_K}} also shows that a cubic
function can approximate the permeability. Similarly to the previous
report \citep{Maillet:2022}, we use the following formula:
\begin{equation}{
\begin{aligned}
K &= K_{0}  \left(\frac{S-S_\mathrm {r}}{1-S_\mathrm {r}}\right)^{3}.
\end{aligned}
}\end{equation} Because both samples of Thiery et al.~and us are
granular media of spherical particles, we apply the Kozeny-Carman
equation, even though its applicability to nano-porous media is
uncertain. \begin{equation}{
K_{0} = \frac{\epsilon^{3} d^{2}}{180  (1-\epsilon)^{2}}.
}\end{equation}

If we approximate \(S \gg S_\mathrm{r}\) and \(S_\mathrm{r}\ll 1\),
\begin{equation}\phantomsection\label{eq:K}{
\begin{aligned}
K &=  K_{0} S ^{3}\\
  &=\frac{\epsilon ^{3}}{180(1-\epsilon)^{2}}d^{2}S ^{3}.
\end{aligned}
}\end{equation} From Eq.~\ref{eq:Pc} and Eq.~\ref{eq:K}, the following
equation is obtained: \begin{equation}{
\begin{aligned}
\frac{K}{\eta}\frac{d}{d S }P_\mathrm{c}&=
-\frac{\epsilon^{3}\Gamma}{180(1-\epsilon)^{2}\alpha\eta}
S ^{3} d. \\
\end{aligned}
}\end{equation}

In conclusion, we can successfully confirm Eq.~\ref{eq:DL2} from the
experimental data and obtain the following equation:
\begin{equation}\phantomsection\label{eq:Dw}{
\widetilde{D}_\mathrm{L}^{0}=\frac{\epsilon^{3}}{180\alpha (1-\epsilon)^{2}}.
}\end{equation} By substituting \(\alpha\sim 0.36\) and
\(\epsilon \sim 0.49\) into Eq.~\ref{eq:Dw}, we get
\(\widetilde{D}_\mathrm{L}^{0}\sim 6.9\times 10^{-3}\).

\section{Appendix C: Nomenclature}

\begin{equation}{
\begin{array}{clcl}
\text{Greek variables} \\
\hline
\textbf{Symbol} & \textbf{Description} & \textbf{unit} & \textbf{First Appearance} \\
\hline
\alpha & \text{Parameters related to capillary pressure} & ND &  \text{Eq. }\ref{eq:Pc} \\
\gamma & := \epsilon/\epsilon_\mathrm{m}& ND & \text{Fig. }\ref{fig:SwvsdD} \\
\Gamma & \text{Surface energy of water} & \text{J/m}^{2} & \text{Eq. }\ref{eq:capillaryrise} \\
\delta & \text{Thickness of the boundary layer} & \text{m} & \text{Eq. }\ref{eq:V}\\
\epsilon & \text{Porosity} & ND & \text{Eq. }\ref{eq:bottom}\\
\epsilon_\mathrm{m} & \text{Microscopic porosity} & ND & \text{Table I }\\
\epsilon_\mathrm{non} & \text{porosity of the
corresponding "non-hi(18) sample"} & ND & \text{Eq. }\ref{eq:non}\\
\eta &  \text{Viscosity of water} & \text{Pa}\cdot\text{s} & \text{Eq. }\ref{eq:DL1}\\
\theta & \text{Contact angle of water} & \text{rad} & \text{Eq. }\ref{eq:capillaryrise} \\
\rho & \text{Density of water} & \text{kg/m}^{3} & \text{Eq. }\ref{eq:capillaryrise}\\
\rho_{0} & \text{Density of water vapor} & \text{kg/m}^{3} & \text{Eq. }\ref{eq:vapor} \\
\sigma & :=P_{0}^{\frac{-1}{n+1}}S   & ND & \text{Eq. }\ref{eq:model1}\\
\sigma^{*} & \sigma \text{ at the transition} & ND & \text{Eq. }\ref{eq:sol1}\\
\tau & \text{Drying time constant} & \text{sec} & \text{Eq. }\ref{eq:exponential}\\
\tilde{\tau} & \text{Fitting parameter} & ND & \text{Eq. }\ref{eq:exponential2}\\
\tau_0 & :=\frac{\epsilon H}{V_{0}}P_{0}^{\frac{1}{n+1}} & \text{sec} & \text{Eq. }\ref{eq:model1}\\
\tau_\mathrm{g} & \text{Diffusion tortuosity} & ND & \text{Eq. }\ref{eq:V} \\
\hline
&  \text{"ND" means non-dimentional}    \\
\end{array}
\nonumber
}\end{equation}

\begin{equation}{
\begin{array}{clcl}
\text{Latin variables} \\
\hline
\textbf{Symbol} & \textbf{Description} & \textbf{unit} & \textbf{First Appearance} \\
\hline
a & \text{Bed area} & \text{m}^{2} & \text{Fig. }\ref{fig:gainen} \\
A & \text{Fitting parameter} & ND & \text{Eq. }\ref{eq:exponential} \\
A' & \text{Fitting parameter} & ND & \text{Eq. }\ref{eq:exponential2} \\
C & \text{Fitting parameter} & ND & \text{Eq. }\ref{eq:exponential} \\
C' & \text{Fitting parameter} & ND & \text{Eq. }\ref{eq:exponential2} \\
d & \text{Diameter of primary particles} & \text{m} & \text{Eq. }\ref{eq:thiery} \\
D & \text{Diameter of secondary particles} & \text{m} & \text{Fig. }\ref{fig:photo} \\
\langle D_\mathrm{L} \rangle_\mathrm{w} & \text{Spatially average of } D_\mathrm{L} \text{ in wet region} & \text{m}^{2}/\text{sec} & \text{Eq. }\ref{eq:enough}\\
D_\mathrm{Lb} & D_\mathrm{L} \text{at the boundary} & \text{m}^{2}/\text{sec} & \text{Eq. }\ref{eq:enough}\\
D_\mathrm{L} & :=\frac{K}{\eta}\frac{dP_\mathrm{c}}{dS } & \text{m}^{2}/\text{sec} & \text{Eq. }\ref{eq:DL1}\\
D_\mathrm{L}^{0} & :=D_\mathrm{L} S ^{-n} & \text{m}^{2}/\text{sec} &  \text{Eq. }\ref{eq:DL2} \\
D_\mathrm{Lo} &  D_\mathrm{L} \text{at the bottom} & \text{m}^{2}/\text{sec} & \text{Eq. }\ref{eq:enough}\\
D_\mathrm{V} & \text{diffusion coefficient of vapor} & \text{m}^{2}/\text{sec} & \text{Eq. }\ref{eq:vapor} \\
F(z) & \text{Approximate function of water saturation} & ND & \text{Eq. }\ref{eq:parabola}\\
h & \text{Thickness of the dry region} & \text{m} & \text{Fig. }\ref{fig:gainen} \\
\tilde{h} & := h/H & ND & \text{Eq. }\ref{eq:model1} \\
H & \text{Thickness of granular bed} & \text{m} & \text{Fig. }\ref{fig:gainen} \\
i & \text{Integer dammy variable} & ND & \text{Eq. }\ref{eq:hetero}\\
K & \text{Permeability of wet region} & \text{m}^{2} & \text{Eq. }\ref{eq:DL1} \\
m & \text{A real constant revealed to be 6.6 } & ND & \text{Eq. }\ref{eq:critical} \\
M & \text{Number of divisions} & ND & \text{Eq. }\ref{eq:risan1} \\
n & \text{A power index revealed to be three } & ND & \text{Eq. }\ref{eq:DL2} \\
N & :=\frac{H\tau_\mathrm{g}}{\delta \epsilon} & ND & \text{Eq. }\ref{eq:V}\\
P & :=\frac{V(H-h)}{\epsilon D_\mathrm{L}^{0}} & ND & \text{Eq. }\ref{eq:kinjikai} \\
P_{0} & :=\frac{V_{0}H}{\epsilon D_\mathrm{L}^{0}} & ND & \text{Eq. }\ref{eq:model1} \\
\widetilde{P_{0}} & := \frac{\eta V_{0}H}{\epsilon \Gamma d} & ND  & \text{Eq. }\ref{eq:trans_power} \\
P_\mathrm{c} & \text{Capillary pressure in wet region} & \text{Pa} & \text{Eq. }\ref{eq:DL1} \\
S & \text{Water saturation} & ND & \text{Eq. }\ref{eq:DL1} \\
\langle S\rangle & S \text{ spatially averaged in entire volume} & ND &\text{Eq. }\ref{eq:Smean} \\
\langle S \rangle^{*} & \langle S\rangle \text{at the transition} & ND & \text{Eq. }\ref{eq:thiery}\\
\langle S\rangle_\mathrm{w} & \text{Spatially averaged water saturation in the wet region} & ND & \text{Eq. }\ref{eq:parabola} \\
S_\mathrm{b} & \text{Approximation of} S \text{ at the boundary} & ND & \text{Eq. }\ref{eq:Sb}\\
S_\mathrm{o} & \text{Approximation of} S \text{ at }z=0 & ND & \text{Eq. }\ref{eq:parabola}\\
S_\mathrm{r} & \text{saturation degree corresponds to the residual water} & ND & \text{Fig. }\ref{fig:Sw_K}\\
t & \text{time} & \text{sec} & \text{Eq. }\ref{eq:DL1}\\
\tilde{t} & :=t/\tau_{0} \text{ (Dimensionless time)} & ND & \text{Eq. }\ref{eq:model1}\\
V & \text{drying rate as the equivalent velocity of liquid water} & \text{m/sec} & \text{Eq. }\ref{eq:h}\\
V_{0} & \text{the value of $V$ in CRP} & \text{m/sec} & \text{Eq. }\ref{eq:model1}\\
z & \text{Upward vertical axis} & \text{m} & \text{Eq. }\ref{eq:continuation} \\
\tilde{z} & :=z/H\text{ (Dimentionless vertical axis)} & ND & \text{Eq. }\ref{eq:model1}\\
\hline
&  \text{"ND" means non-dimentional}    \\
\end{array}
\nonumber
}\end{equation}

%

\end{document}